\documentclass[sigconf,nonacm,screen]{acmart}
\usepackage{booktabs}
\usepackage{needspace}
\usepackage{xurl} 
\usepackage{tikz}
\usetikzlibrary{arrows.meta,positioning}

\newcommand{\mAltAlphaMonth}{29}
\newcommand{\mAltBetaMonth}{3.2}
\newcommand{\mColdNaiveMax}{7.7}
\newcommand{\mColdNaiveMin}{5.7}
\newcommand{\mColdOurs}{6.2}
\newcommand{\mColdOursCached}{5.4}
\newcommand{\mColdOvFourStreams}{6.7}
\newcommand{\mColdPt}{7.7}
\newcommand{\mCorpusCharMax}{523}
\newcommand{\mCorpusCharMin}{13}
\newcommand{\mCorpusChunks}{17}
\newcommand{\mCorpusN}{16}
\newcommand{\mCorpusTokMax}{476}
\newcommand{\mCorpusTokMin}{13}
\newcommand{\mDayIdleSharePt}{95}
\newcommand{\mDayOurs}{0.38}
\newcommand{\mDayOvSized}{3.24}
\newcommand{\mDayPt}{2.41}
\newcommand{\mEffOrtDefault}{0.89}
\newcommand{\mEffOrtSizedB}{2.37}
\newcommand{\mEffOurs}{2.71}
\newcommand{\mEffOursB}{3.26}
\newcommand{\mEffOvDefault}{1.35}
\newcommand{\mEffOvFourStreams}{2.31}
\newcommand{\mEffOvSized}{2.71}
\newcommand{\mEffOvSizedB}{3.17}
\newcommand{\mEffPtDefaultB}{0.69}
\newcommand{\mFirstOurs}{0.94}
\newcommand{\mFirstOvDefault}{1.26}
\newcommand{\mFirstOvEight}{0.69}
\newcommand{\mFirstOvFourStreams}{0.56}
\newcommand{\mGainOverOrt}{3.1}
\newcommand{\mGainOverPt}{4.1}
\newcommand{\mIdleOrt}{4.0}
\newcommand{\mIdleOursFloor}{1.33}
\newcommand{\mIdleOursFloorB}{1.13}
\newcommand{\mIdleOursFloorMb}{1331}
\newcommand{\mIdleOursPre}{12.1}
\newcommand{\mIdleOvDefault}{5.2}
\newcommand{\mIdleOvSizedEight}{12.1}
\newcommand{\mIdlePt}{8.7}
\newcommand{\mIdlePtB}{11.0}
\newcommand{\mIdlePtMb}{8659}
\newcommand{\mIdleReductionX}{6.5}
\newcommand{\mLoneShortOurs}{493}
\newcommand{\mLoneShortOv}{134}
\newcommand{\mOvDefaultLossPct}{50}
\newcommand{\mPadAudioX}{3.1}
\newcommand{\mPadShortTok}{36}
\newcommand{\mPadTo}{128}
\newcommand{\mPocketColdFirstEver}{13.2}
\newcommand{\mPocketColdWarm}{5.5}
\newcommand{\mPocketIdle}{1234}
\newcommand{\mPocketPostIdle}{1.12}
\newcommand{\mRestoreMax}{2.3}
\newcommand{\mRestoreMaxB}{1.2}
\newcommand{\mRestoreMed}{2.2}
\newcommand{\mRestoreMedB}{1.1}
\newcommand{\mRestoreMin}{2.1}
\newcommand{\mRestoreMinB}{1.0}
\newcommand{\mRestoreN}{10}
\newcommand{\mRestoreNB}{10}
\newcommand{\mStaticDeltaPct}{-3.3}
\newcommand{\mThrPlateau}{38}
\newcommand{\mUsdAudioHOurs}{0.0153}
\newcommand{\mUsdAudioHPt}{0.0631}
\newcommand{\mUsdIdleMonthPt}{69}
\newcommand{\mUtilPlateauEight}{14.9}
\newcommand{\mUtilPlateauSixteen}{14.6}

\begin{document}

\title{Pushing CPU Speech Synthesis to the Wall:\\
Extreme Inference Tuning under Serverless Architecture \& Billing}

\author{Pakorn Nathong}
\affiliation{%
  \institution{Paxa Labs}
  \country{} 
}
\email{research@paxalabs.com}
\author{Kunat Pipatanakul}
\affiliation{%
  \institution{Paxa Labs}
  \country{}
}

\renewcommand{\shortauthors}{Technical Report}
\renewcommand{\shorttitle}{Technical Report}

\authorsaddresses{}

\begin{abstract}
Instance-billed serverless platforms charge not only for active computation
but also for CPU and memory held across the lifetime of a warm instance,
turning idle inference state into a direct serving cost. We present
billing-aware serving for neural text-to-speech (TTS) on serverless CPUs,
where the optimization objective is therefore the CPU-seconds and GB-seconds
accumulated over an instance lifetime rather than throughput or latency in
isolation. Existing inference runtimes are poorly matched to this setting
in two ways: machine-wide per-request parallelism causes CPU contention when
independent utterances execute concurrently, while warm instances retain
gigabytes of billable inference state and page-cache memory during idle
periods. Scaling to zero removes the idle-memory charge, but replaces it
with a multi-second process cold start.

We address these two costs with request-sized concurrent inference and a
reclaimable instance lifecycle. The former bounds per-request CPU
parallelism so independent requests efficiently share a fixed execution
budget. The latter releases compiled inference state and billable page-cache
memory after a quiet period while retaining the server process and compile
cache for faster restoration.

On Kokoro-82M, the resulting system sustains
\mEffOurs{} audio-seconds per CPU-second, compared with
\mEffOrtDefault{} for ONNX~Runtime defaults, and reduces cost per
audio-hour from \$\mUsdAudioHPt{} for PyTorch to
\$\mUsdAudioHOurs{}, a \mGainOverPt$\times$ reduction relative to
PyTorch and \mGainOverOrt$\times$ relative to ONNX~Runtime defaults.
It reduces idle billed memory from \mIdlePt{}\,GB for PyTorch to
\mIdleOursFloor{}\,GB, a \mIdleReductionX$\times$ reduction, and
restores a released instance to first audio in \mRestoreMed{}\,s,
against \mColdPt{}\,s for a PyTorch process cold start, while
preserving reference output. Under bursty traffic, request sizing alone
can increase total cost because idle memory dominates the bill; the
reclaimable lifecycle is what converts busy-time inference efficiency
into lower serverless cost. These results show that serverless CPU
inference should be organized around the billed instance lifetime rather
than the always-on execution model assumed by conventional inference
runtimes.
\end{abstract}

\maketitle

\section{Introduction}
\label{sec:intro}

Small neural text-to-speech (TTS) models are attractive CPU workloads:
per-request computation is modest, and serverless containers offer
fine-grained capacity with scale-to-zero. The inference runtimes used to
serve these models, however, are designed primarily for always-on
deployment. This mismatch matters on instance-billed platforms, which
charge for CPU and memory over the container's lifetime, including idle
periods. At the worked-example rates used throughout this paper,
$\alpha=\$30$ per vCPU-month and $\beta=\$8$ per
GB-month,\footnote{Round figures of the same order as current public list
prices: AWS Fargate charges about \$\mAltAlphaMonth{} per vCPU-month and
\$\mAltBetaMonth{} per GB-month on demand~\cite{aws2026fargate}, while
Fly.io charges about \$5 per GB-month for added
memory~\cite{fly2026pricing}. Our conclusions compare configurations and
do not depend on these exact rates.}
a PyTorch server retaining \mIdlePt{}\,GB while idle costs roughly
\$\mUsdIdleMonthPt{} per month before serving any traffic. Scaling to zero
avoids that charge, but the next request waits \mColdPt{}\,s for first
audio.

The optimization objective is therefore not throughput or latency in
isolation, but the CPU-seconds and GB-seconds accumulated over the
instance lifetime. We call serving that explicitly targets this objective
\emph{billing-aware serving}. Our thesis is that real-time TTS on
instance-billed serverless CPUs should be organized around this billed
lifetime rather than around the always-on execution model assumed by
current inference runtimes.

Our measurements reveal two structural problems
(Section~\ref{sec:motivation}). First, concurrent inference is poorly
matched to framework threading defaults: requests cannot be continuously
batched, yet each inference call is given machine-wide parallelism, so
concurrent requests contend for the same CPUs. Second, the usual instance
lifecycle exposes an unattractive choice between warm and cold: staying
warm retains several gigabytes of billable inference state, while scaling
to zero pays a multi-second process restart. 

We address the two structural problems directly
(Section~\ref{sec:design}). \emph{Request-sized concurrent inference}
bounds per-request CPU parallelism so that independent requests share a
fixed execution budget efficiently. A \emph{reclaimable instance
lifecycle} adds a released state between warm-idle and process cold:
after a quiet period, the server relinquishes inference state and
billable page-cache memory while retaining the process and compile cache
needed for a faster restore.

\begin{figure}[t]
  \centering
  \includegraphics[width=0.9\linewidth]{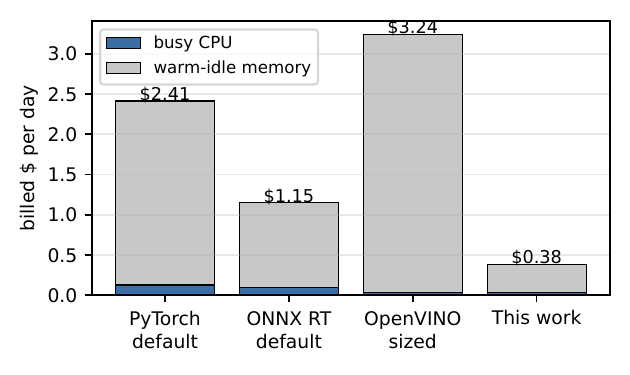}
  \caption{Billed cost for one day of bursty traffic: two hours of
  synthesized audio delivered in eight equal bursts. Request sizing
  improves busy efficiency but leaves a large warm-idle footprint; the
  reclaimable lifecycle removes the idle-memory component that otherwise
  dominates the bill.}
  \label{fig:money}
\end{figure}

On Kokoro-82M, the resulting system reaches \mEffOurs{}
audio-seconds per CPU-second and reduces busy cost to
\$\mUsdAudioHOurs{} per audio-hour, \mGainOverPt{}$\times$ below PyTorch
defaults and \mGainOverOrt{}$\times$ below ONNX~Runtime defaults. It
reduces warm-idle memory from \mIdlePt{}\,GB to
\mIdleOursFloor{}\,GB and restores a released instance in
\mRestoreMed{}\,s, compared with \mColdPt{}\,s for a PyTorch process
cold start. First-audio latency remains no worse than the PyTorch and
ONNX~Runtime defaults. More importantly, under bursty traffic, request
sizing alone can increase total cost because idle memory dominates the
bill; the reclaimable lifecycle is what converts the busy-time efficiency
gain into lower serverless cost (Figure~\ref{fig:money}).
Together, the two mechanisms capture the dominant measured cost reduction.
After these serving-path changes, the lower-level optimizations we test
either measure within repeat-run variability or fail output equivalence.

\Needspace*{6\baselineskip}
\noindent\textbf{Contributions.}
\begin{itemize}
\item We characterize neural TTS serving under instance-billed
serverless CPUs and identify two structural inefficiencies: machine-wide
per-request parallelism under concurrency and billable inference state
retained during idle periods (Section~\ref{sec:motivation}).

\item We introduce request-sized concurrent inference and a reclaimable
instance lifecycle that directly address the busy and idle components of
the bill (Section~\ref{sec:design}).

\item We evaluate both mechanisms independently and end to end,
characterize the remaining optimization headroom after these changes, and
show that the design conclusions transfer across CPU architectures and to
an independently developed TTS system (Section~\ref{sec:results}).
\end{itemize}

\section{Motivation and System Observations}
\label{sec:motivation}

Serverless deployment changes the serving objective. Rather than
optimizing throughput or latency in isolation, an instance-billed server
must minimize the CPU and memory accumulated over its lifetime. We
characterize how existing inference runtimes behave under this objective
and derive the two requirements that drive our design.

\subsection{Serverless Billing Changes the Objective}
\label{sec:billing}

Serverless platforms either charge resources only while requests are
active or meter the container over its entire lifetime. We study the
latter, \emph{instance-billed} regime, where CPU-seconds and GB-seconds
are charged at rates $\alpha$ and $\beta$. A warm instance therefore
continues to incur memory cost while idle, whereas scaling to zero removes
that cost but makes the next request pay a cold start.

The memory meter also includes the cgroup's page cache, so model files
retained by the kernel remain billable until reclaimed. All dollar results
use the worked-example rates from Section~\ref{sec:intro}. Under this
billing model, reducing idle memory is a cost optimization, not merely a
capacity optimization.

\subsection{Concurrency and Request Parallelism}
\label{sec:obs-threads}

The exported graph synthesizes one utterance per call, so independent
requests execute concurrently rather than through continuous batching.
Padding variable-length requests without masking changes the synthesized
output (Section~\ref{sec:negative}), and the model lacks the token-by-token
decoding structure used by continuous-batching LLM servers.

This execution model interacts poorly with framework defaults.
ONNX~Runtime and OpenVINO size each request's thread pool to the
machine's physical cores~\cite{ort2024threading,openvino2025hints}, so
concurrent requests compete for the same CPUs. At four clients, their
default configurations achieve only \mEffOrtDefault{} and
\mEffOvDefault{} audio-seconds per CPU-second, respectively.
Efficient serving therefore requires allocating CPU parallelism
\emph{across} concurrent requests rather than giving each request the
whole machine.

\subsection{Idle Memory and Instance Lifecycle}
\label{sec:obs-idle}
\label{sec:obs-lifecycle}

Inference runtimes retain allocator state, caches, and weight mappings
after traffic stops. Ten minutes after the last response, resident memory
remains at \mIdlePt{}\,GB for PyTorch, \mIdleOrt{}\,GB for
ONNX~Runtime, and \mIdleOvDefault{}\,GB for OpenVINO. On bursty traffic,
this warm-idle memory dominates the bill for configurations without a
release mechanism (Figure~\ref{fig:money}).

Scaling to zero avoids that memory charge but exposes the opposite cost:
a fresh process takes \mColdNaiveMin{}--\mColdNaiveMax{}\,s to reach
first audio across the evaluated backends. Thus the usual two choices,
fully warm or process cold, are both unattractive for bursty interactive
traffic. Serverless TTS needs an intermediate state that relinquishes
most billed inference memory while restoring substantially faster than a
process restart.

\section{Design}
\label{sec:design}

\begin{figure}[!ht]
\centering
\begin{tikzpicture}[
  font=\scriptsize,
  box/.style={draw, rounded corners=1pt, minimum height=5.5mm, inner sep=2pt, align=center},
  st/.style={draw, rounded corners=2pt, minimum height=6mm, minimum width=16mm, inner sep=2pt, align=center, fill=gray!12},
  arr/.style={-{Latex[length=1.6mm]}, thick},
  lab/.style={font=\scriptsize, align=center, inner sep=1pt}]
\node[lab, anchor=west] at (-0.7,0.6) {\textbf{Request path}};
\node[box] (req) at (0,0) {HTTP\\request};
\node[box, right=3.5mm of req] (g2p) {G2P\\front end};
\node[box, right=3.5mm of g2p] (pool) {inference-request pool\\8 streams $\times$ 2 threads};
\node[box, right=3.5mm of pool] (out) {chunked\\audio};
\draw[arr] (req) -- (g2p);
\draw[arr] (g2p) -- (pool);
\draw[arr] (pool) -- (out);

\node[lab, anchor=west] at (-0.7,-1.0) {\textbf{Instance lifecycle}};
\node[st] (busy) at (0.8,-2.2) {busy};
\node[st, right=12mm of busy] (idle) {warm-idle};
\node[st, right=12mm of idle] (rel) {released};
\draw[arr] (busy) -- node[lab, above=3.6mm] {last response} (idle);
\draw[arr] (idle) -- node[lab, above=3.6mm] {quiet window:\\release state} (rel);
\draw[arr] (idle.south) to[bend left=35]
  node[lab, below] {request} (busy.south);
\draw[arr] (rel.south) to[bend left=32]
  node[lab, below, pos=0.55] {request: restore from cache} (busy.south west);
\end{tikzpicture}
\caption{Serving design. Top: concurrent requests use bounded
per-request parallelism from a fixed inference pool. Bottom: after an
idle period, the server releases inference state and restores it from
the compile cache on the next request.}
\label{fig:arch}
\end{figure}
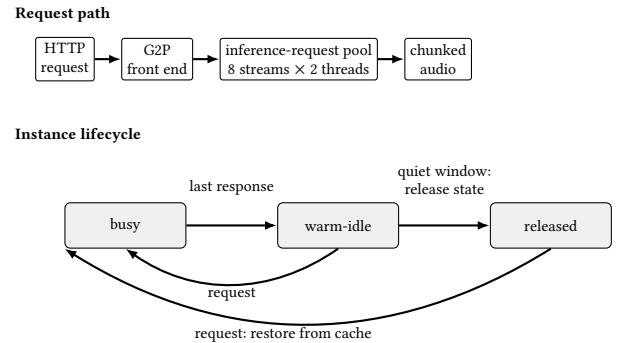

Section~\ref{sec:motivation} identifies two serving problems. Concurrent
requests contend when each inference call is given machine-wide
parallelism, while warm-idle instances retain billable inference state.
We address them with two mechanisms: request-sized concurrent inference
(Section~\ref{sec:sched}) and a reclaimable instance lifecycle
(Section~\ref{sec:lifecycle}). Figure~\ref{fig:arch} summarizes both.

\subsection{Request-Sized Concurrent Inference}
\label{sec:sched}

Because requests cannot be continuously batched
(Section~\ref{sec:obs-threads}), serving parallelism comes from executing
independent requests concurrently. Giving each request a machine-wide
thread pool oversubscribes the CPU, so we instead bound per-request
parallelism and serve requests through a fixed pool of inference streams.

Our evaluated configuration uses eight streams with two threads each on
a 32-vCPU container. Requests therefore execute independently while
sharing a fixed CPU budget rather than competing through overlapping
machine-wide thread pools. Synthesis remains chunked so that the server
can return the first audio chunk before completing the utterance.
Section~\ref{sec:res-sizing} evaluates the throughput and latency
tradeoff of this design.

\subsection{Reclaimable Instance Lifecycle}
\label{sec:lifecycle}

To avoid paying continuously for warm inference state, we add a
\emph{released} state between warm-idle and process cold
(Figure~\ref{fig:arch}). After a quiet period, the server frees the
compiled model and request-pool state, returns unused allocator memory,
and evicts its compile-cache files from the billable page cache. The
server process and on-disk compile cache remain available.

On the next request, the server reconstructs inference state from the
compile cache rather than restarting and recompiling from scratch. Release
and request handling are synchronized so that inference state is never
discarded while a request is active. The resulting state gives up most
warm-idle memory while retaining a shorter restore path than scale-to-zero.
Section~\ref{sec:res-lifecycle} evaluates this tradeoff.

\section{Implementation}
\label{sec:implementation}

Our implementation is approximately 1900 lines of Python and uses
OpenVINO 2026.3. A common serving shim receives text over HTTP, applies
the reference text-to-phoneme front end, runs inference, and streams
audio in chunks. The same request path is used for PyTorch eager,
ONNX~Runtime, and OpenVINO, so all measurements cover the complete path
from text input to audio output.

The inference backend, thread and stream counts, compile-cache directory,
and idle timeout are configurable at start-up. The OpenVINO backend
implements the reclaimable lifecycle of Section~\ref{sec:lifecycle} by
freeing inference state, trimming allocator memory, writing back and
evicting the compile-cache files from the page cache, and reconstructing
the compiled model from its on-disk cache. Release and request handling share a lock
to prevent state reclamation during active inference.

We implement no custom kernels. The artifact, is publicly available.\footnote{\url{https://github.com/paxalabs/cpu-tts-to-the-wall}}
\section{Experimental Setup}
\label{sec:setup}

\begin{table*}[t]
\caption{End-to-end serving of Kokoro-82M on the primary 32-vCPU
Emerald-Rapids host. Efficiency and first-audio latency are measured at
four concurrent clients; idle memory is measured 600\,s after the last
response unless noted; cold start is a fresh process start to first
audio; and cost uses $(\alpha,\beta)$. The sized OpenVINO idle value is
measured after 30\,s, before any release. Restore is the time to first
audio for the first request after a release. Best measured values are in
\textbf{bold}.}
\label{tab:e2e}
\centering\small
\begin{tabular}{lcccccc}
\toprule
 & audio-s/CPU-s$\,\uparrow$
 & first audio p50/p90$\,\downarrow$
 & warm-idle MB$\,\downarrow$
 & cold start s$\,\downarrow$
 & restore s$\,\downarrow$
 & \$/audio-h$\,\downarrow$ \\
\midrule
PyTorch eager
  & 0.66 & 2.46/8.0 & 8659 & 7.7 & -- & 0.0631 \\
ONNX Runtime default
  & 0.89 & 0.94/4.9 & 3969 & 5.8 & -- & 0.0468 \\
ONNX Runtime sized
  & 1.36 & 2.36/8.4 & 3958 & \textbf{5.7} & -- & 0.0306 \\
OpenVINO default
  & 1.35 & 1.26/\textbf{2.6} & 5250 & 6.9 & -- & 0.0309 \\
OpenVINO sized
  & \textbf{2.71} & \textbf{0.69}/4.2 & 12097 & 7.4 & -- & 0.0154 \\
This work
  & \textbf{2.71} & 0.94/4.2 & \textbf{1331} & 6.2 & 2.2 & \textbf{0.0153} \\
\bottomrule
\end{tabular}
\par\smallskip
{\scriptsize\raggedright\emph{Note:} ``--'' indicates that restore is
not supported.\par}
\end{table*}

\subsection{Model and Workload}
\label{sec:setup-model}

We evaluate Kokoro-82M~\cite{hexgrad2025kokoro}, a
StyleTTS2-family~\cite{li2023styletts2} pipeline with an
iSTFTNet-style~\cite{kaneko2022istftnet} convolutional decoder producing
24\,kHz audio, using the fp32 community ONNX export. All backends use
the same text-to-phoneme front end, so measurements cover the complete
request path from text to audio. We verify output equivalence against
the reference backend on every corpus chunk.

The workload contains \mCorpusN{} utterances
(\mCorpusChunks{} model chunks) spanning
\mCorpusCharMin{}--\mCorpusCharMax{} characters and
\mCorpusTokMin{}--\mCorpusTokMax{} tokens per chunk. The corpus is fixed
across runs and chosen to cover input lengths. Under closed-loop load,
clients repeatedly cycle through this corpus, so faster configurations
complete more requests during the same measurement window.

\subsection{Environment and Measurement}
\label{sec:setup-hw}
\label{sec:setup-load}

Primary experiments run in a 32-vCPU container on an Intel Xeon
Platinum 8581C (Emerald Rapids); transfer experiments use a 48-vCPU
container on an AMD EPYC 9655 (Zen~5). Both run under shared tenancy.
The software stack is Linux 6.18, Python 3.12.14, PyTorch 2.6.0 (CPU),
ONNX~Runtime 1.20.1, and OpenVINO 2026.3.

Load is generated inside the serving container using a closed loop:
each of $c$ clients submits its next request when the previous one
completes. We exclude ramp-up and measure server CPU time over the same
steady-state window. In-process generation avoids an external network
bottleneck becoming the limiting resource. Key configurations are
repeated at least twice. We retain a candidate optimization only if it preserves reference
output on every corpus chunk and improves performance beyond repeat-run
variability. This criterion excludes apparent gains from a single run and
configurations that change what the model computes.

\subsection{Configurations and Metrics}
\label{sec:setup-baselines}

We evaluate five baselines and our complete system. PyTorch eager uses
its defaults. ONNX~Runtime is evaluated with default threading and with
four intra-op threads per request. OpenVINO is evaluated with its
defaults and with the request sizing of Section~\ref{sec:sched}; sizing
uses a total budget of sixteen threads across four or eight inference
streams. The end-to-end sized baseline in Table~\ref{tab:e2e} uses
eight streams.

For the bursty-cost analysis in Figure~\ref{fig:money}, we use a
synthetic 24-hour trace with two audio-hours of demand divided evenly
across eight bursts. The container is not scaled to zero between bursts:
baseline configurations remain warm-idle, while our system may enter the
released state after its 45\,s quiet window.

Our complete system uses eight two-thread OpenVINO streams, a compile
cache, and the reclaimable lifecycle of Section~\ref{sec:lifecycle}
with a 45\,s quiet window.

We report sustained efficiency as audio-seconds produced per CPU-second,
and first-audio latency as p50 and p90, both at four concurrent clients.
We additionally measure resident memory after idle, process cold-start
latency, restore latency from the released state, and cost per audio-hour
under the worked-example rates $(\alpha,\beta)$ from
Section~\ref{sec:intro}. We define restore latency as the time to first
audio for the first request after a release, reported as the median and
range over \mRestoreN{} release-and-restore cycles
(Section~\ref{sec:res-lifecycle}).

\section{Results}
\label{sec:results}

We first measure the end-to-end effect of the complete system. We then
isolate the effect of request-sized concurrent inference on busy-time
efficiency and the effect of the reclaimable instance lifecycle on idle
memory and restore latency. We next measure the remaining optimization
headroom and finally test whether the design conclusions transfer across
CPU and model architectures.

\subsection{End-to-End Serving Cost}
\label{sec:e2e}

Our complete system reduces both the busy and idle components of the
serverless bill. Table~\ref{tab:e2e} shows that it reaches
\mEffOurs{} audio-seconds per CPU-second and reduces busy cost to
\$\mUsdAudioHOurs{} per audio-hour, \mGainOverPt{}$\times$ below
PyTorch defaults and \mGainOverOrt{}$\times$ below ONNX~Runtime
defaults. At four clients, median first-audio latency increases from
\mFirstOvEight{} to \mFirstOurs{}\,s relative to sized OpenVINO but
remains no worse than the PyTorch and ONNX~Runtime defaults.

The reclaimable instance lifecycle reduces billed memory after the quiet
period from \mIdlePtMb{}\,MB for PyTorch to \mIdleOursFloorMb{}\,MB for
our complete system. A request arriving after release reaches first audio
in \mRestoreMed{}\,s, compared with \mColdOurs{}\,s for a process cold
start of the same configuration. The released state therefore reduces
idle memory without making the next burst pay a process cold start.

Figure~\ref{fig:money} shows why both mechanisms are necessary under
bursty traffic. Request sizing alone improves busy-time efficiency but
leaves \mIdleOvSizedEight{}\,GB resident, increasing the daily bill to
\$\mDayOvSized{}, compared with \$\mDayPt{} for PyTorch. Adding the
reclaimable instance lifecycle reduces the complete-system bill to
\$\mDayOurs{}. Because warm-idle memory accounts for
\mDayIdleSharePt\% of the PyTorch bill, the lifecycle is what converts
the busy-time efficiency gain into lower total cost.

\subsection{Request-Sized Concurrent Inference}
\label{sec:res-sizing}

Table~\ref{tab:ablation} isolates the busy-time effect of request sizing.
Starting from OpenVINO defaults, a sixteen-thread budget across four
streams increases efficiency from \mEffOvDefault{} to
\mEffOvFourStreams{} audio-seconds per CPU-second and reduces median
first-audio latency at four clients from \mFirstOvDefault{} to
\mFirstOvFourStreams{}\,s. Expanding the pool to eight streams raises
efficiency further to \mEffOvSized{}. Leaving OpenVINO at its defaults
therefore costs \mOvDefaultLossPct\% of sustained efficiency at this
load.

\begin{figure}[!ht]
\centering
\includegraphics[width=\linewidth]{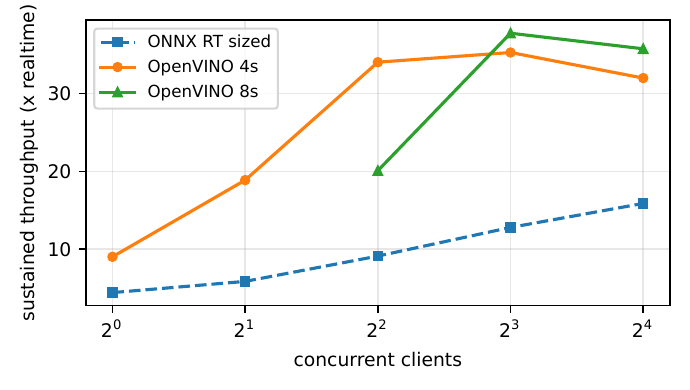}
\caption{Sustained throughput with increasing concurrency. Sized
configurations scale to eight clients and then plateau; additional
requests primarily queue.}
\label{fig:capacity}
\end{figure}

Figure~\ref{fig:capacity} shows that sized configurations scale to eight
concurrent clients and then plateau near \mThrPlateau{}$\times$ realtime.
Across the plateau, CPU use remains between \mUtilPlateauSixteen{} and
\mUtilPlateauEight{} of 32 logical CPUs, so additional requests
primarily increase queueing rather than throughput.

Request sizing trades isolated-request latency for more efficient
concurrent execution. In a separate single-request test, a short
utterance reaches first audio in \mLoneShortOv{}\,ms with OpenVINO
defaults and \mLoneShortOurs{}\,ms with request sizing, remaining below
half a second in both cases. At four clients, Table~\ref{tab:ablation}
shows that request sizing improves both sustained efficiency and median
first-audio latency. The latency tradeoff therefore appears when a single
request could otherwise use the whole machine.

\begin{table}[!ht]
\caption{Mechanism ablation on the primary host. Efficiency and first
audio are measured at four clients; idle memory is measured after
600\,s, except for the eight-stream configuration measured after 30\,s
before release. Cold start is a process start to first audio, from a
cached compile where a cache exists. Restore is the first-audio time of
the first request after a release, reported as the median of \mRestoreN{}
release-and-restore cycles.}
\label{tab:ablation}
\centering\scriptsize
\setlength{\tabcolsep}{1.5pt}
\begin{tabular}{@{}lcccccc@{}}
\toprule
Configuration
& eff.$\,\uparrow$
& \shortstack{first p50\\s$\,\downarrow$}
& \shortstack{idle\\MB$\,\downarrow$}
& \shortstack{cold\\s$\,\downarrow$}
& \shortstack{restore\\s$\,\downarrow$}
& \shortstack{\$/aud-h\\$\downarrow$} \\
\midrule
OpenVINO defaults
  & 1.35 & 1.26 & 5250 & 6.9 & -- & 0.0309 \\
+ sized threads, 4 streams
  & 2.31 & 0.56 & 8684 & 6.7 & -- & 0.0180 \\
+ compile cache
  & 2.33 & 0.64 & 7490 & 5.4 & -- & 0.0179 \\
+ 8 streams
  & 2.71 & 0.69 & 12097 & 7.1 & -- & 0.0154 \\
+ idle release (this work)
  & 2.71 & 0.94 & 1331 & 6.2 & 2.2 & 0.0153 \\
\bottomrule
\end{tabular}
\par\smallskip
{\scriptsize\raggedright\emph{Note:} ``--'' indicates that restore is
not supported.\par}
\end{table}

\subsection{Reclaimable Instance Lifecycle}
\label{sec:res-lifecycle}

\begin{figure}[!ht]
\centering
\includegraphics[width=\linewidth]{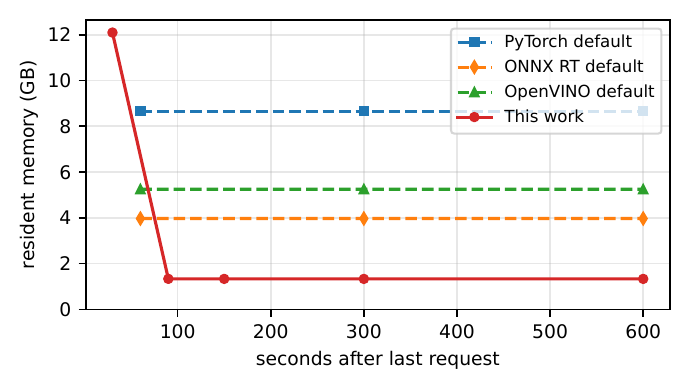}
\caption{Resident memory after the last request. Framework baselines
retain their warm-idle footprint, while the reclaimable lifecycle reaches
its released-state floor after the quiet period.}
\label{fig:idle}
\end{figure}

The released state is useful only if it reduces warm-idle memory while
restoring faster than a process restart. Figure~\ref{fig:idle} shows the
memory effect. The framework baselines retain their warm-idle footprints
throughout the ten-minute window, while our server drops from
\mIdleOursPre{}\,GB to \mIdleOursFloor{}\,GB after the 45\,s quiet period.

The final three rows of Table~\ref{tab:ablation} separate compile caching,
the eight-stream request pool, and release. The compile cache leaves
steady-state efficiency essentially unchanged while reducing process
cold start from \mColdOvFourStreams{} to \mColdOursCached{}\,s.
Increasing the pool to eight streams provides the remaining efficiency
gain but raises warm-idle memory to \mIdleOursPre{}\,GB. Releasing
inference state then reduces billed memory to \mIdleOursFloorMb{}\,MB
without changing sustained efficiency.

Restore measurements show the corresponding latency tradeoff. A request
arriving after release reaches first audio in \mRestoreMed{}\,s (range
\mRestoreMin{}--\mRestoreMax{}\,s over \mRestoreN{} cycles), against
\mColdOurs{}\,s for a process cold start of the same configuration. The
released state therefore occupies the intended middle point between
warm-idle and process cold: it relinquishes most billed inference memory
and reaches first audio nearly three times faster than restarting the
process.

\subsection{Optimization Headroom}
\label{sec:negative}

After addressing the concurrency and idle-state mismatches, we profile
the graph to bound the remaining optimization headroom. Execution time
concentrates in \texttt{Sin}, \texttt{Reorder}, convolutions, and
\texttt{STFT}, while the fusable element-wise work targeted by
static-shape compilation and kernel fusion is small. This profile suggests
limited end-to-end headroom from these techniques for this graph.

We test static-shape compilation, which can enable operator fusion and
static memory planning, and padding-based bucketing for the variable input
lengths that fixed-shape execution requires. Under the acceptance
criterion of Section~\ref{sec:setup-load}, static-shape compilation
changes per-call time by \mStaticDeltaPct\%, within repeat-run
variability. Padding without masking fails output equivalence: padding a
\mPadShortTok-token utterance to \mPadTo{} tokens produces
\mPadAudioX$\times$ as much audio because the model synthesizes the
padding. We therefore retain neither optimization.

Taken together, these results bound the optimization headroom measured in
this study. After fixing the major serving-path inefficiencies, the tested
graph and compiler optimizations provide little additional end-to-end
improvement for this model and hardware. We leave broader investigation
of conventional CPU optimizations across other models and hardware to
future work.

\subsection{Transfer}
\label{sec:transfer}

On the Zen~5 host, the same design reaches \mEffOursB{} audio-seconds per
CPU-second against \mEffPtDefaultB{} for PyTorch defaults, and releases to
\mIdleOursFloorB{}\,GB against the \mIdlePtB{}\,GB PyTorch holds while
idle. Its median restore latency is \mRestoreMedB{}\,s (range
\mRestoreMinB{}--\mRestoreMaxB{}\,s over \mRestoreNB{} cycles). The
backend gap narrows on this architecture: sized ONNX~Runtime reaches
\mEffOrtSizedB{}, compared with \mEffOvSizedB{} for sized OpenVINO. The backend advantage is therefore
architecture-dependent, while the benefits of request-sized concurrent
inference and the reclaimable instance lifecycle remain.

We separately test the lifecycle observation on
pocket-tts~\cite{kyutai2025calm}, an independently developed
100M-parameter continuous audio language model served by its own
PyTorch-based HTTP server. These measurements use a separate
32-vCPU Emerald-Rapids-class container and are not compared directly
with Table~\ref{tab:e2e}.

\begin{figure}[!ht]
\centering
\includegraphics[width=\linewidth]{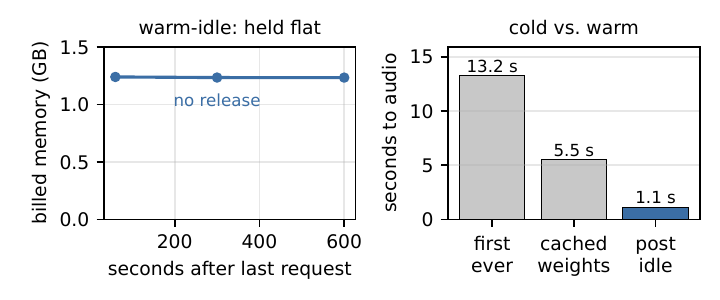}
\caption{pocket-tts under the same lifecycle analysis. Left: billed
memory remains resident after traffic stops. Right: serving exposes
warm and process-cold states, with no intermediate release state.}
\label{fig:pocket}
\end{figure}

Figure~\ref{fig:pocket} shows the same lifecycle structure. pocket-tts
reaches first audio in \mPocketColdFirstEver{}\,s on its first start and
\mPocketColdWarm{}\,s with cached weights. After serving, it retains
\mPocketIdle{}\,MB throughout the ten-minute idle window, while a
post-idle request returns in \mPocketPostIdle{}\,s because no state was
released. Thus an independently developed model exhibits the same
warm-idle versus cold-start tradeoff that motivates the reclaimable
lifecycle.

\section{Related Work}
\label{sec:related}

\noindent\textbf{StyleTTS2 inference.} StyleTTS~2 provides HiFi-GAN- and
iSTFTNet-based waveform decoders~\cite{kong2020hifigan,kaneko2022istftnet,li2023styletts2}.
Relative to matched HiFi-GAN variants, iSTFTNet replaces output-side neural
layers with magnitude and phase prediction followed by an inverse STFT,
reducing learned computation. Streaming systems instead reduce first-output
latency through incremental generation~\cite{ellinas2020streamingtts}.
These techniques optimize one utterance; we target cross-request concurrency
and idle state while preserving the Kokoro-82M graph, weights, and output.

\noindent\textbf{CPU inference stacks.} PyTorch~\cite{paszke2019pytorch},
ONNX~Runtime~\cite{ort2024threading}, OpenVINO~\cite{openvino2025hints},
and oneDNN~\cite{onednn2026} provide optimized kernels and thread or stream
controls. We build atop them, sizing CPU across concurrent TTS requests and
managing billable state between bursts.

\noindent\textbf{Serverless inference.} Workload studies motivate
keep-warm policies~\cite{shahrad2020serverless}; MArk combines batching,
autoscaling, and serverless capacity~\cite{zhang2019mark}. They manage cold
starts or fleet capacity; ours releases model and page-cache state inside a
warm TTS container, preserving faster restore than process cold.

\section{Limitations}
\label{sec:limitations}

This study covers one model family (StyleTTS2/Kokoro, evaluated on
Kokoro-82M), one runtime family (OpenVINO for the approach; PyTorch and
ONNX~Runtime as baselines), two CPU architectures, and a synthetic
length-mixed trace. The negative results of Section~\ref{sec:negative} are scoped
accordingly. Static-shape compilation, for example, should help on graphs
dominated by fusable element-wise operators; this export is not one, and
we did not evaluate quantization, SMT, or core pinning.
The transfer check of Section~\ref{sec:transfer} covers only the billing and
lifecycle claims.

\section{Conclusion}
\label{sec:conclusion}

Serverless CPU platforms bill an integral that current inference
frameworks were not designed to minimize. Restructuring the serving path
around that integral, with request-sized concurrent inference and a
reclaimable instance lifecycle that releases billed memory including page
cache and restores from compile caches in seconds, reduces the cost of
StyleTTS2-family TTS by \mGainOverPt$\times$ per audio-hour relative to
PyTorch defaults, and by far more on bursty traffic where warm-idle memory
dominates the bill. Beyond this configuration, added concurrency stops
converting into throughput and further serving-side tuning measures at
or below noise, so further improvement is determined by the model rather
than by the serving path.


\bibliographystyle{ACM-Reference-Format}
\bibliography{refs}

\end{document}